# High-resolution urban air pollution and thermal comfort mapping: an application of drive-by sensing platform for smart city services

Hui Zhong [a,c,#], Hongliang Lu [a,c,#], Ting Gan[d], Yonghong Liu[d,*], Xinhu Zheng [a,b,c,*]

[a] *Intelligent Transportation Thrust, Systems Hub, The Hong Kong University of Science and Technology (Guangzhou), Guangzhou, China;*

*b Guangdong Provincial Key Lab of Integrated Communication, Sensing and Computation for Ubiquitous Internet of Things, Guangzhou, China;*

[c] *School of Intelligent Systems Engineering, Sun Yat-sen University, Guangzhou, 510006, China;*

**Abstract:** Air pollutant exposure exhibits significant spatial and temporal variability, with localized hotspots, particularly in traffic microenvironments, posing health risks to commuters. Although widely used for air quality assessment, fixed-site monitoring stations are limited by sparse distribution, high costs, and maintenance needs, making them less effective in capturing on-road pollution levels. This study utilizes a fleet of 314 taxis equipped with sensors to measure $NO_2$, $PM_{2.5}$, and $PM_{10}$ concentrations and identify high-exposure hotspots. The findings reveal disparities between mobile and stationary measurements, map the spatiotemporal exposure patterns, and highlight local hotspots. These results demonstrate the potential of mobile monitoring to provide fine-scale, on-road air pollution assessments, offering valuable insights for policymakers to design targeted interventions and protect public health, particularly for sensitive populations.

**Keywords**: on-road air pollution; exposure risk; mobile monitoring; spatiotemporal; emerging technology

*Corresponding author

*E-mail address:* liuyh3@mail.sysu.edu.cn. (Yonghong Liu)

*E-mail address:* xinhuzheng@hkust-gz.edu.cn. (Xinhu Zheng)

# Equal contribution

# 1. Introduction

Air pollution poses a significant threat to public health, contributing to approximately 7 million premature deaths worldwide each year due to high-risk exposure to pollutants (Organization, 2019). Previous studies have demonstrated that non-communicable diseases, including cardiovascular disease, lung cancer, and ischemic heart disease, are responsible for 70% of deaths attributable to air pollution (Wang et al., 2016; Landrigan, 2017). There are six typical pollutants for regulatory air pollution monitoring, i.e., particles with aerodynamic diameters under 2.5 and 10 μm ($PM_{2.5}$ and $PM_{10}$), nitrogen oxides (NO and $NO_2$), carbon monoxide (CO) and sulfur dioxide ($SO_2$) (Suh et al., 2000; Landrigan et al., 2018). Short-term exposure (from one hour to several days) to PM and $NO_2$ with increased risks of respiratory, and cardiovascular mortality, as well as hospital admissions and emergency department visits (Schwarz et al., 2024; Yu et al., 2024).

Traffic-related air pollution has drawn increasing attention, with vehicle emissions becoming a major source of urban pollutants (Künzli et al., 2000; Li et al., 2013; Zhong et al., 2023). Moreover, air pollutant concentrations often exist spatial and temporal heterogeneity over short distances ( 0.01–1 km) within urban areas (Karner et al., 2010; Zhu et al., 2002). Some traffic microenvironments (TMEs), such as street canyons, bus stops, intersections, and roadsides, frequently exhibit pollutant levels that far exceed those of the surrounding ambient environment. Research shows that concentrations of traffic-related pollutants like PM and $NO_2$ in these areas can reach up to ten times the background levels (Westerdahl et al., 2005; Sexton et al., 2004; Liu et al., 2021). Additionally, various active transportation modes, such as walking, cycling, scootering, and running, often occur within these TMEs. Therefore, although individuals may spend only 7%–10% of their day commuting, this limited exposure can disproportionately contribute to 12%–30% of their total daily inhalation of pollutants due to high concentrations in these commuting environments (Shen and Gao, 2019; Che et al., 2020). In China, for instance, exposure to traffic-related PM2.5 alone is estimated to result in 1.6 million premature deaths annually (Tian et al., 2018).

Traditionally, fixed-site monitors are widely used to assess air pollution exposure levels. However, these instruments are often expensive and sparsely distributed, with intensive maintenance and calibration protocols (Apte et al., 2017; Yu et al., 2022). Moreover, their typical placement away from roadsides further limits their effectiveness in capturing on-road pollution levels. To achieve finer spatial resolution, some studies combine emission modeling tools with air dispersion models to simulate pollutant concentrations at virtual receptors (Vallamsundar et al., 2016; Deng et al., 2020; Zhong et al., 2023). However, these models heavily depend on emission inventories and often fail to account for unanticipated pollution sources. Besides, satellite-based remote sensing provides broad-scale exposure concentrations but with a relatively coarse spatial resolution (> 1–10 km), which is insufficient for detecting fine-scale gradients and local high-exposure hotspots (Buchholz et al., 2021; Kumar, 2010).

Recent advancements in low-cost, portable environmental sensors have offered promising opportunities for assessing urban air pollution exposure. Vehicle-based monitoring has emerged as a widely adopted approach, enabling the deployment of air quality sensors on mobile platforms such as garbage trucks, public buses, and taxis to facilitate real-time measurement of on-road pollutant levels (Apte et al., 2017; Wu et al., 2020; DeSouza et al., 2020; Batur et al., 2022). Compared to traditional monitoring and modeling methods, vehicle-based sensors can capture fine-scale spatiotemporal variations influenced by factors like weather and traffic, providing extensive coverage and allowing for semi-continuous measurement areas. Thus, high-resolution air pollution mapping and hotspot identification become feasible, overcoming the location constraints of stationary monitors. For example,

Apte et al. (2017) used Google Street View vehicles to map $NO_2$ concentrations at a fine scale, while DeSouza et al. (2020) employed low-cost optical particle counters on trash trucks to detect $PM_{2.5}$ hotspots in Cambridge, MA. However, such studies have primarily been conducted in developed countries, using limited numbers of vehicles and generally focusing on single pollutants. A comprehensive assessment is lacking to systematically characterize exposure concentrations of multiple pollutants in spatiotemporal with an extensive fleet of sensor-equipped vehicles.

In this study, we deployed 314 taxis equipped with sensors for multiple air pollutants, including $NO_2$, $PM_{2.5}$, and $PM_{10}$, to monitor on-road air pollution exposure in Guangzhou, China. Unlike trash trucks, non-scheduled taxis operate 24 hours a day and frequently traverse busy streets with intensive human activity, where pollutant concentrations tend to be elevated. Therefore, this study aimed to achieve three objectives: (1) evaluate the exposure disparities of multiple air pollutants between mobile monitors and stationary measurements; (2) analyze the spatiotemporal patterns of exposure concentrations; and (3) recognize local high-exposure hotspots. The findings underscore the potential of mobile monitoring to provide accessible, fine-scale urban air pollution data, offering valuable insights and supporting air pollution control efforts in densely populated developing countries.

## 2. Materials and Methods

### *2.1 Study design*

This study was conducted in the central urban area of Guangzhou, which has a permanent population of 18.7 million and spans 7,434 km² (Bureau, 2023). The region features a 2,263-km road network classified into primary, secondary, motorway, residential, and tertiary roads. Six fixed air quality monitoring stations, illustrated in Fig. 1, were selected to support the validation and comparison of on-road air pollution measurements (Wu et al., 2020). A fleet of 314 non-scheduled taxis was equipped with portable sensors to measure on-road concentrations of $NO_2$, $PM_{2.5}$, $PM_{10}$, ambient temperature, and vehicle trajectories. Over the study period, from February 1 to March 31, 2023, approximately 5 million data records were collected, spanning 1,440 hours and covering 562.6 km of roadways. The operational flexibility of taxis, including frequent traversals of high-activity urban roads, ensured comprehensive spatiotemporal coverage. On average, each 50-meter road segment was sampled for 59 days, yielding 448 data points, enabling a granular analysis of spatial distribution patterns and exposure variability across diverse TMEs. Simultaneously, air quality data from fixed monitoring stations were utilized to evaluate potential exposure disparities in mobile measurements and to enhance the reliability of the collected data through calibration.

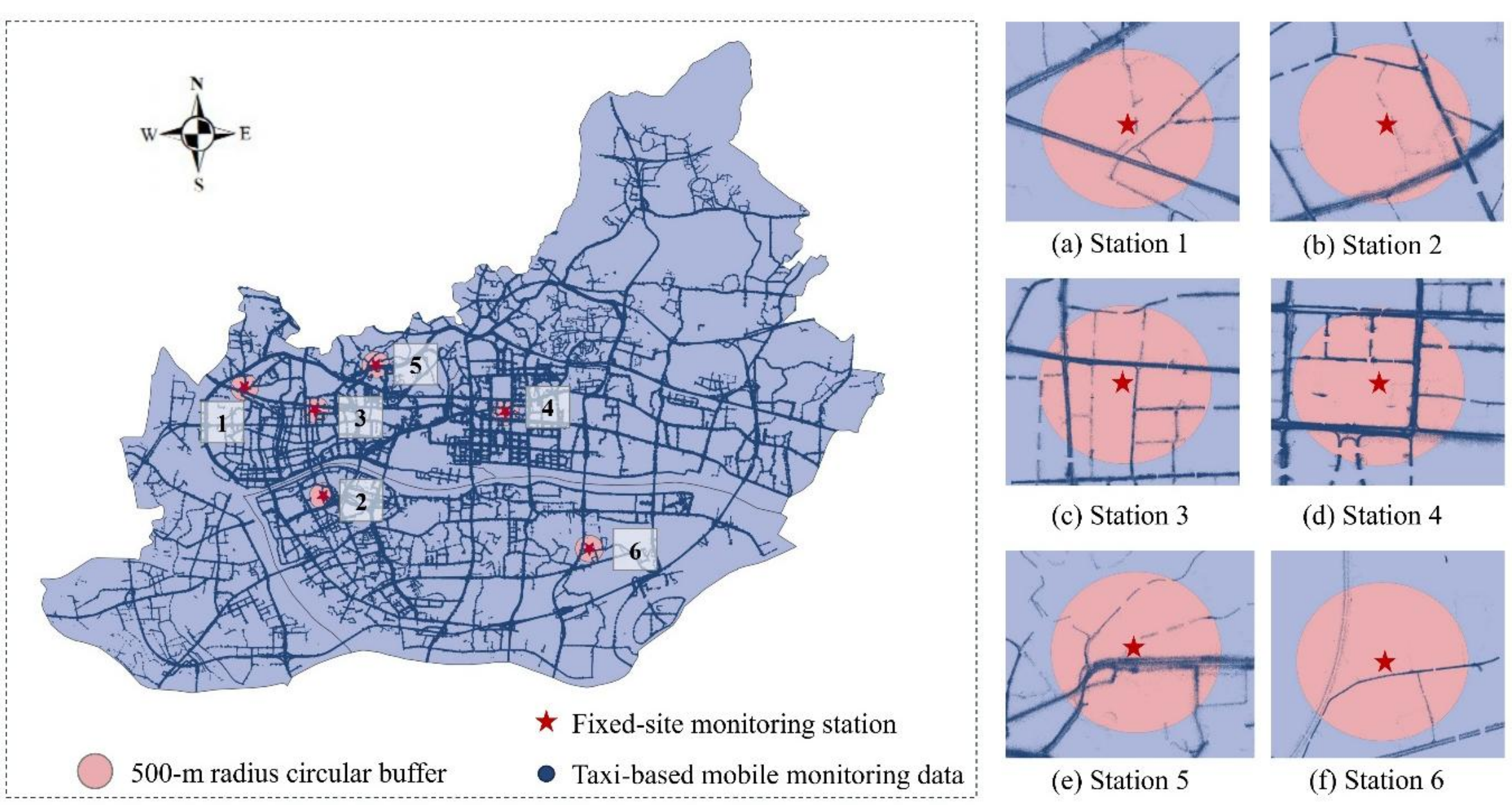

Fig.1 The mobile monitoring data and fixed-site station distribution in the study area.

*2.2. Instruments and quality assurance*

To capture high-resolution on-road air pollution and environmental data, a sensor suite was installed within the roof-mounted lightbox of each taxi. This system measured ambient temperature, humidity, $PM_{2.5}$, $PM_{10}$, and $NO_2$ concentrations, along with corresponding GPS coordinates. Deployment configurations are detailed in Fig. S1. The monitoring instruments complied with Chinese standards for ambient air quality continuous automated monitoring systems (HJ6, 2013a, b). To ensure measurement accuracy, a Thermo 146i NO$x$ analyzer was employed to calibrate $NO_2$ readings and validate the calibration curves. For $PM_{2.5}$ and $PM_{10}$, calibration and validation were conducted by comparing mobile sensor data with simultaneous roadside measurements from fixed air quality monitoring stations in Guangzhou. This approach ensured that the mobile system accurately captured traffic-related particulate matter concentrations and reflected spatial variability trends within the region. To further verify the reliability of the instruments, sensors installed on three taxis were subjected to parallel testing to assess inter-device consistency. $PM_{2.5}$ and $PM_{10}$ concentrations in ambient air were continuously measured over 24 hours, with at least ten sampling groups per test. The relative deviation across devices was maintained below 15%, satisfying precision requirements. Additionally, fully assembled sensor systems underwent aging tests on a dedicated test rack, operating continuously for over 24 hours to confirm stable performance. Table 1 summarizes each monitoring instrument's technical specifications, including measurement ranges and error margins.

Table 1 Parameters of instruments

| Instrument Accuracy | Measurement | Accuracy | Response Time (s) | Resolution |
| --- | --- | --- | --- | --- |
| ASAir AHT25 | Temperature | ±0.3℃ | <5.0 | 0.01 ℃ |
| | Humidity | ±2% | <8.0 | 0.024% RH |
| Alphasense NO2-B43F | $NO_2$ | ±15% | <80.0 | 0.1 ppb |
| Plantower PMS5003T | $PM_{2.5}$<br>$PM_{10}$ | Maximum of ±10%<br>and ±10 μg/m3 | <1.0 | 0.1 $\mu g/m^3$ |
| GPS | coordinates | ±10 m | - | 0.00001 |

*2.3. Data reduction*

The extensive dataset generated from frequent and long-term measurements posed significant challenges for spatiotemporal analysis using conventional statistical approaches. To address this, a data reduction algorithm was implemented to condense approximately 5 million instantaneous observations into median hourly concentration estimates for individual 50-m road segments (Apte et al., 2017). The first step involved a "snapping" process, where each GPS-recorded measurement was assigned to the nearest 50-m road segment according to the nearest neighbor algorithm (Taunk et al., 2019). It was a suitable choice to balance spatial resolution—avoiding excessive variability seen in smaller segments—and sample size, helping reduce the impact of outliers and GPS errors that may arise in fragmented datasets. Therefore, each 50-m segment was assigned a single concentration estimate. This implies that it is feasible for each segment to serve as a unit for characterizing localized hotspots and analyzing traffic microenvironments (TMEs).

This study included approximately 26,052 road segments, with data collected by 314 vehicles. Among these segments, 62% had data coverage spanning 50–60 distinct days, while 86% had at least 12 hours of observations. Apte et al. (2017) suggested that the median concentration was a robust measure of central tendency, particularly suited for skewed datasets. As shown in Fig. S2, the distributions of on-road $NO_2$, $PM_{2.5}$, and $PM_{10}$ concentrations exhibited slight skewness, with median values consistently lower than their corresponding means. This pattern highlights the advantages of using median concentrations that are less influenced by extreme values to provide a more reliable representation of typical pollution levels across the study area.

### *2.4. Evaluation of exposure disparities and hotspots identification*

Generally, on-road exposure concentrations were relatively higher than situated air quality monitoring stations, especially during rush hours (Huang et al., 2023). To evaluate the exposure differences between on-road and ambient air pollution, comparisons were made with hourly measurements from six national fixed-site monitoring stations within the study area. A 500-m radius circular buffer surrounded each station, and mobile monitoring data falling within these buffers were extracted for comparison analysis (Wu et al., 2020). Due to the relationship between taxi riding trajectories and human activity intensity, each station buffer's sample sizes vary, ranging from about 40,000 to 420,000 records from stations 1 to 6. Five metrics were introduced to compare the consistency for $NO_2$, $PM_{2.5}$, and $PM_{10}$ exposure concentrations between mobile and fixed-site measurements, including fractional bias (FB), normalized mean square error (NMSE), geometric variance (VG), correlation coefficient (R), and the fraction of predictions within a factor of two of the observations (FAC2). Also, the relative errors (EF, $ER = \left|\frac{C_m - C_f}{C_f}\right| \times 100\%$) was calculated to evaluate the exposure bias.

Van den Bossche et al. (2015) demonstrated that mobile monitoring is a powerful tool for mapping urban air quality at high spatial resolution, revealing spatial variability that stationary monitors cannot capture. In this study, dense road network sampling enabled the identification of localized high-exposure hotspots, which are essential for understanding urban air quality and associated health risks. To identify high-exposure hotspots, we developed a hotspot identification algorithm that recognizes high-risk areas based on elevated exposure concentrations. Exposure levels for $PM_{2.5}$, $NO_2$, and $PM_{10}$ were determined using the air quality index (AQI) standards from the U.S. EPA, with specific thresholds outlined in Table S1. The algorithm's *get_level* function maps the measured concentrations to corresponding exposure levels according to these predefined thresholds.

We introduced two key metrics to ensure data reliability and reduce uncertainty: valid monitoring time (VMT) and valid monitoring segment (VMS). VMT represents the effective duration during which exposure concentration $C_j$ at time $T_j$ was sampled on a given segment. If consecutive sampling intervals were less than 30 minutes apart, the exposure concentration at $T_j$ was considered valid. A segment was classified as a VMS only if its total valid monitoring time exceeded 10 days (Apte et al., 2017; Wu et al., 2020; DeSouza et al., 2020). These criteria helped exclude segments with insufficient or intermittent data, ensuring that identified hotspots were robust and representative of sustained high-exposure conditions.

**Algorithm 1** Hotspot Identification Algorithm

1: **Input:** $R$: Set of road segments (each 50 m), *pollution_con:* Exposure concentrations of $PM_{2.5}$, $PM_{10}$, and $NO_2$
2: **Output:** Exposure level for each pollutant on each road segment
3: **Initialize:**
4: $R$: Road set with $N$ elements
5: $Rtime$: Road Level Time Set with $N$ elements
6: $Rlevel$: Road Level Set with $N$ elements
7: Let pollutants $P[k]$ represent $PM_{2.5}$ ($k = 1$), $NO_2$ ($k = 2$), $PM_{10}$ ($k = 3$)
8: Define ***get_level*** ($value$, $k$)
9: **for** $k = 1$ to 3 **do**
10: **if** $value$ is within range for $P[k]$ **then**
11: Return level: 1 = Good, 2 = Moderate, 3 = Unhealthy for Sensitive Groups, 4 = Unhealthy
12: **end if**
13: **end for**
14: **for** $i = 1$ to $N$ **do**
15: Initialize counters: $good[k]$, $moderate[k]$, $unhealthy_sensitive[k]$, $unhealthy[k]$ for $k = 1, 2, 3$
16: **for** $j = 2$ to $N$ **do**
17: **if** $Tj - Tj{-}1 < 1800$ s **then**
18: **for** $k = 1$ to 3 **do**
19: $level \leftarrow get_level\ (P[k]_{j-1}, k)$
20: Update counters based on $level$
21: **end for**
22: **end if**
23: **end for**
24: $Rtimei[k] \leftarrow$ argmax($good[k]$, $moderate[k]$, $unhealthy_sensitive[k]$, $unhealthy[k]$)
25: Determine $Rleveli[k]$: if $Rtimei[k]$ equals *unhealthy_sensitive* or *unhealthy,* mark as "hotspot"; else "not hotspot"
26: **end for**

## 3. Results and conclusion

*3.1. Disparities in exposure measurements between fixed and mobile monitoring*

Fixed-site monitoring stations, typically positioned away from roadways, often fail to capture the elevated exposure concentrations characteristic of on-road environments. To bridge this gap, interpolation and land-use regression have been employed to estimate exposure levels in TMEs (Qi et al., 2022; Cai et al., 2020; Hankey and Marshall, 2015). Thus, it is crucial to quantify the relationship and variations between fixed-site and mobile monitoring data to assess their reliability and applicability for evaluating on-road exposure.

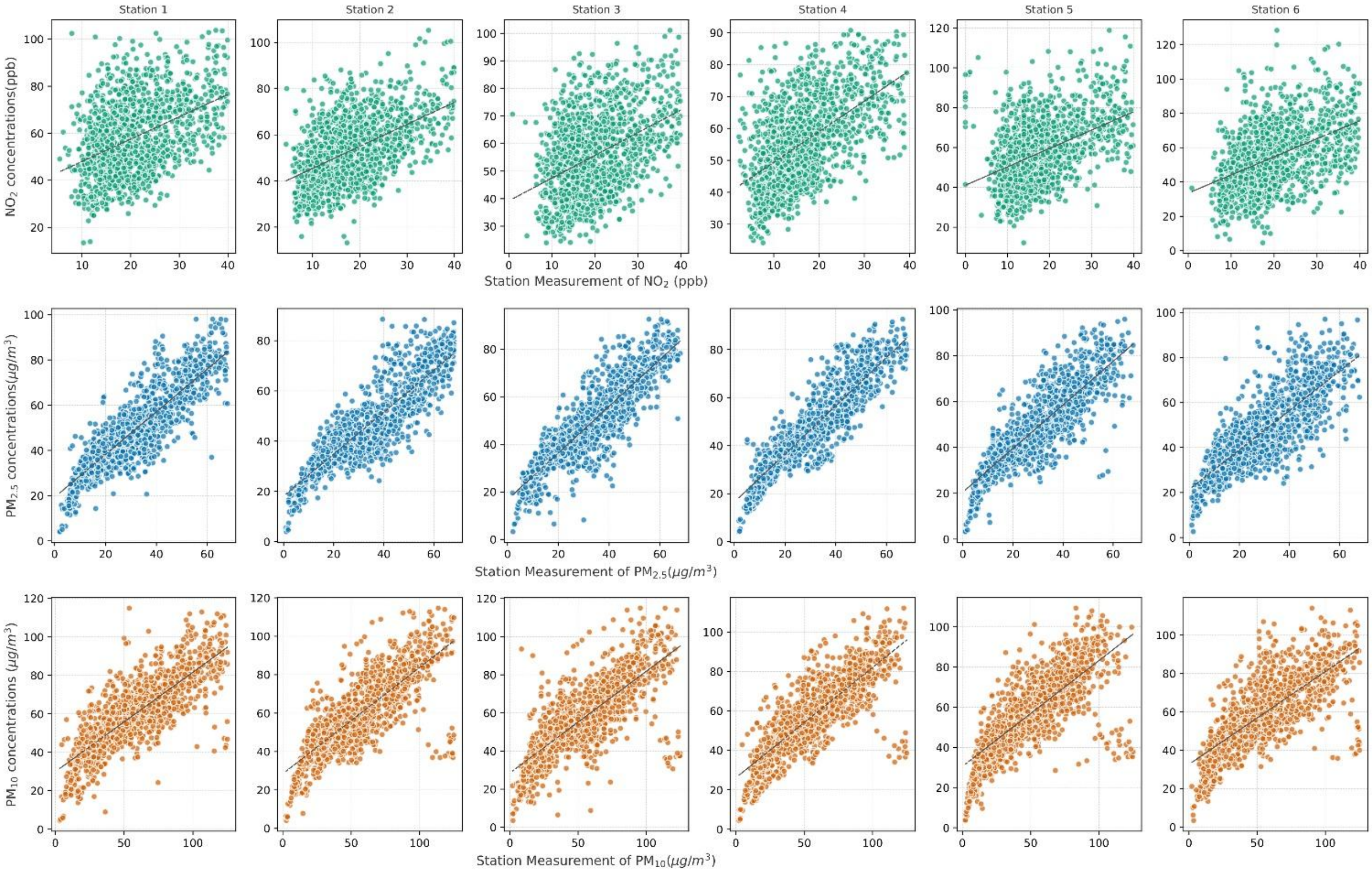

Fig.2 Comparison of $NO_2$, $PM_{2.5}$, and $PM_{10}$ Exposure Concentrations Between Mobile Monitoring and Fixed-Site Stations

Fig.2 compares hourly data from six fixed-site stations and mobile monitoring within a 500-m buffer, comprising 1,440 data points per station. Fig.S3 evaluates their consistency using five performance metrics. PM concentrations showed strong correlations, with coefficients (R) of 0.78 for $PM_{2.5}$ and 0.88 for $PM_{10}$, while $NO_2$ exhibited moderate correlation ($R$ = 0.48). FB and NMSE are commonly used to evaluate the agreement between datasets, while VG and the FAC2 provide insights into absolute differences (Wu et al., 2020). $PM_{2.5}$ and $PM_{10}$ met the thresholds ($-0.5 < FB < 0.5$, NMSE $< 0.5$), indicating strong agreement, while $NO_2$ showed higher values, highlighting notable inconsistencies. For absolute differences, $PM_{2.5}$ and $NO_2$ exhibited VG values exceeding 1.3 and FAC2 below 0.8, indicating notable discrepancies. However, $PM_{10}$ showed smaller differences, with VG and FAC2 values falling within acceptable ranges. These findings suggest that PM exposure concentrations generally align with variation trends observed at fixed stations but differ significantly in absolute values, while $NO_2$ is more strongly influenced by localized traffic emissions.

Fig.3 quantifies the exposure errors between fixed-site and on-road monitoring. On-road data consistently showed higher concentrations, with $NO_2$ exceeding fixed-site levels by 0.6–1.1, $PM_{2.5}$ by 0.6–1.0, and $PM_{10}$ by 0.35–0.45. Within a 1-km resolution (500-m radius) buffer zone, relying solely on fixed-site measurements to estimate exposure concentrations would lead to substantial underestimation. For instance, $PM_{10}$ exposure would be underestimated by at least 35%, while $NO_2$ exposure errors could exceed 100%. These results underscore the limitations of fixed-site monitors in capturing the spatial variability and localized hotspots characteristic, highlighting the critical role of mobile monitoring in accurately assessing urban air pollution exposure (Van den Bossche et al., 2015; Huang et al., 2023).

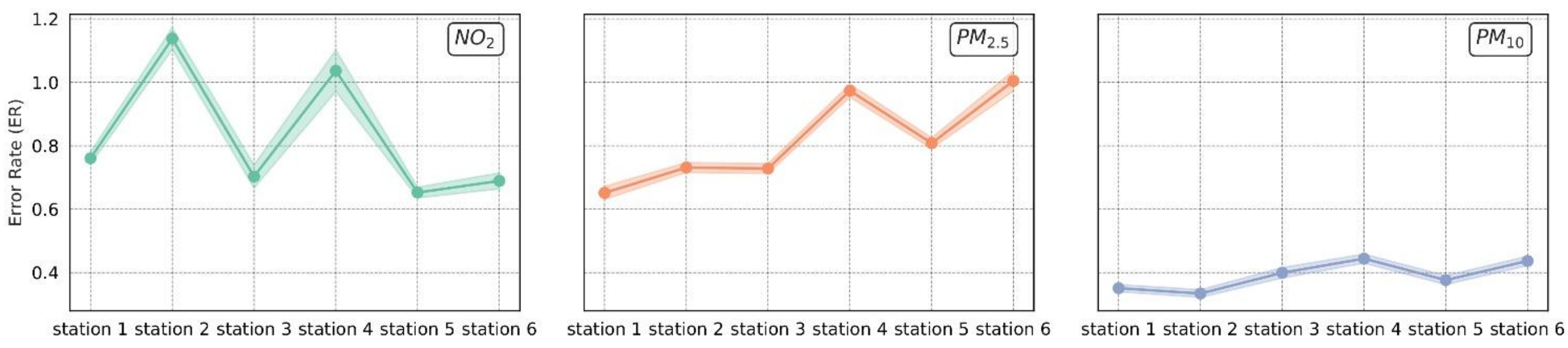

Fig.3 Exposure disparities of $NO_2$, $PM_{2.5}$ and $PM_{10}$ from taxi-based mobile monitoring data and stationary sites

*3.2. Temporal patterns of exposure concentrations*

During the observation period, weather conditions remained stable, with temperatures ranging from 15 to 30°C and relative humidity between 40% and 75%. As transportation is a major contributor to on-road air pollution, the temporal variability of on-road exposure concentrations warrants investigation. Transportation is a significant source of on-road air pollution, exhibiting distinct temporal patterns influenced by traffic flow and atmospheric conditions. Thus, it is essential to capture the variability of exposure on the road.

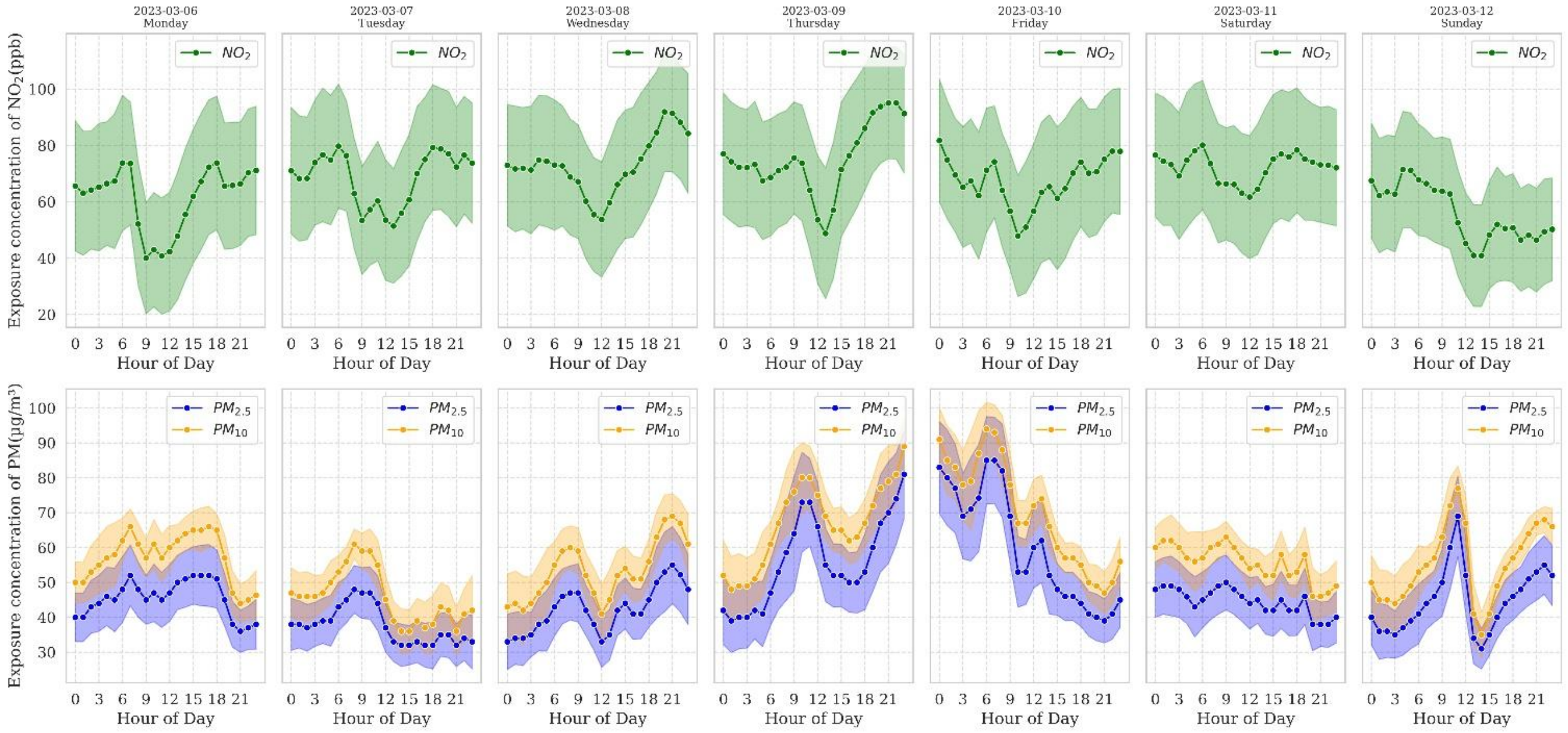

Fig.4 Hourly exposure concentrations for $NO_2$, $PM_{2.5}$ and $PM_{10}$ in a typical week (error bar: 1 SD)

Fig.4 illustrates the average hourly concentrations of $NO_2$, $PM_{2.5}$, and $PM_{10}$ over a typical week, with error bars denoting ±1 standard deviation. $NO_2$ exhibited a consistent daily trend, with concentrations gradually decreasing through the morning, reaching a minimum around noon, and then steadily increasing during the afternoon. In contrast, $PM_{2.5}$ and $PM_{10}$ displayed synchronized variations, with concentrations rising wavily and peaking between Thursday night and early Friday morning. This weekly accumulation effect aligns with findings by Wu et al. (2020), which attribute elevated nighttime pollution levels to reduced atmospheric mixing.

Fig.5 illustrates the hourly exposure concentrations across different road types to explore the impact of vehicle flows. The temporal patterns of $PM_{2.5}$ and $PM_{10}$ generally show a double-peak trend, with morning peaks between 8:00–10:00 and evening peaks from 18:00–20:00. Whereas, $NO_2$ exhibits an inverse diurnal pattern, with a minimum concentration at noon and a peak around 18:00. As highlighted by Kendrick et al. (2015); Qi et al. (2022), NO emitted from vehicle exhaust reacts with $O_3$ to produce $NO_2$ during daylight hours. However, with increasing solar radiation and rising temperatures, $NO_2$ undergoes photolysis, converting back into $O_3$, which explains the midday decline in $NO_2$ levels. In the evening, as solar radiation diminishes, the photolysis of $NO_2$ slows significantly, while the reaction of NO with $O_3$ persists, leading to an accumulation of $NO_2$ and a peak concentration around 18:00. Following this peak, reduced vehicular traffic further decreases NO emissions, resulting in a gradual decline in $NO_2$ levels overnight. These trends align with Fig.S4, which shows corresponding variations in temperature and humidity over 24 hours, further supporting the observed diurnal patterns.

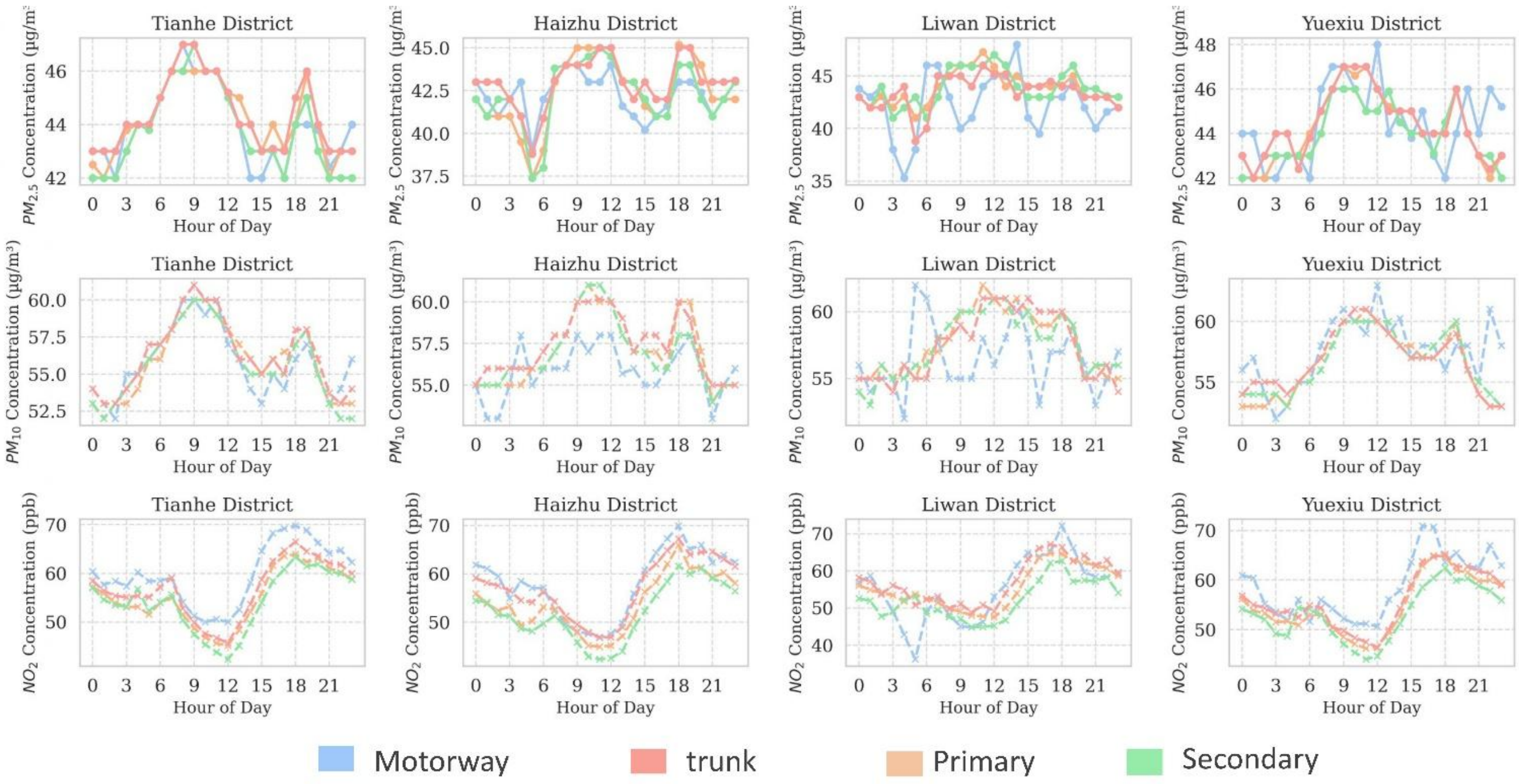

Fig.5 Hourly exposure concentrations for NO2, PM2.5 and PM10 on different road types

### *3.3. High-exposure hotspots identification*

Widely used methods for mapping spatiotemporal exposure levels, such as the natural breaks classification method, often represent concentration levels using varying color intensities (Batur et al., 2022; Huang et al., 2023). Fig.6 demonstrates the spatiotemporal distribution of $PM_{2.5}$, $PM_{10}$, and NO2 exposure concentrations at specific hours. High $NO_2$ concentrations are predominantly observed along motorways and trunk roads during the afternoon. In contrast, $PM_{2.5}$ and $PM_{10}$ exhibit lower concentrations during off-peak hours, with their spatial distribution becoming more distributed and characterized by frequent “red alert” zones during peak periods for the relatively small variation in PM concentrations across these zones, leading to difficulties of directly identifying high-exposure hotspots and evaluating associated health risks.

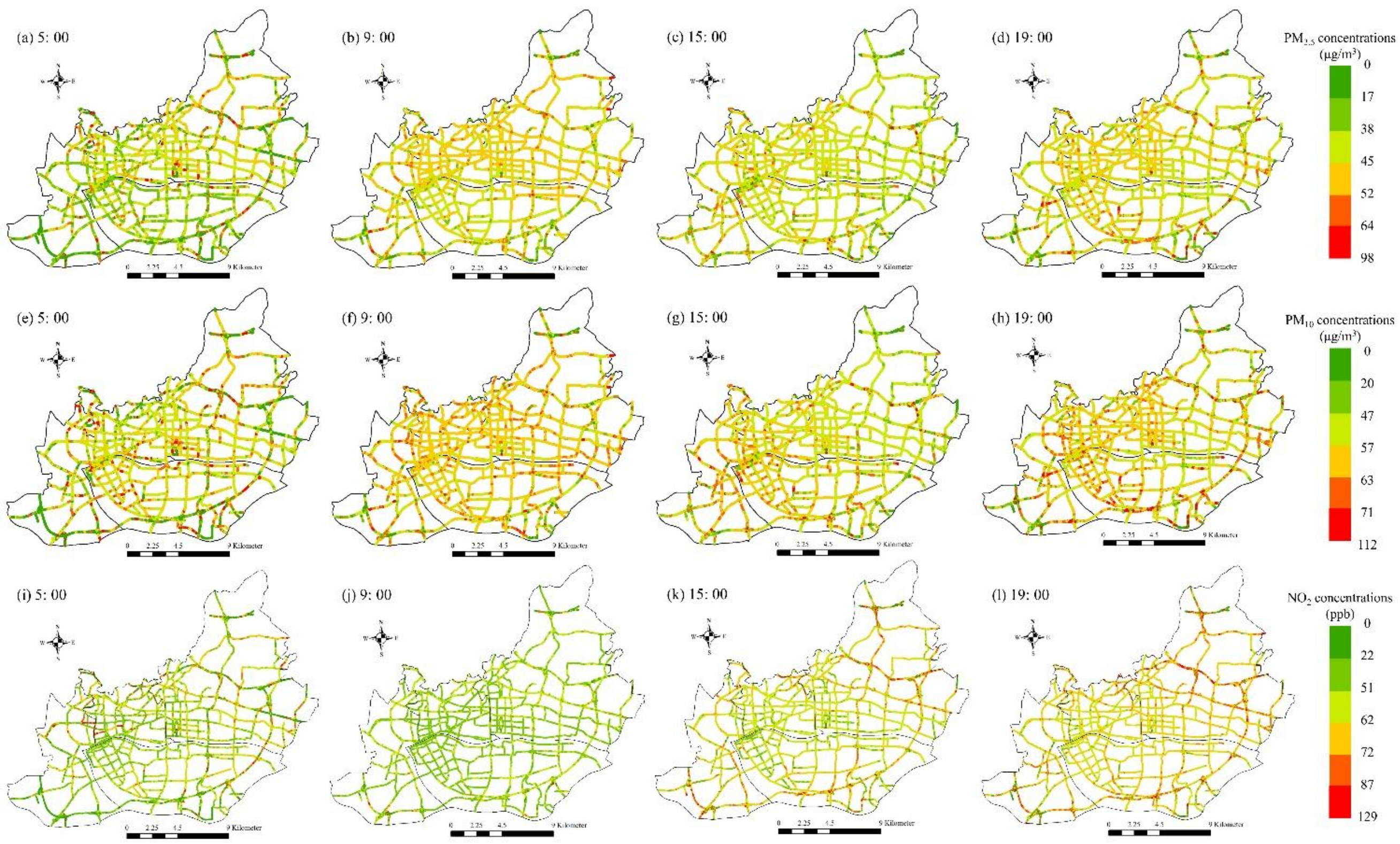

Fig.6 Spatial-temporal exposure pattern of $NO_2$, $PM_{2.5}$ and $PM_{10}$ at special hours

To address this issue, exposure levels are categorized into four distinct classes based on concentration thresholds: 1) Good: Air quality is satisfactory, posing minimal or no health risk; 2) Moderate: Air quality is acceptable, though minor health effects may occur for sensitive individuals; 3)

Unhealthy for Sensitive Groups: Sensitive individuals may experience health risks, while the general population is less likely to be affected; 4) Unhealthy: Elevated risks affect the general population, with sensitive groups experiencing more severe effects.

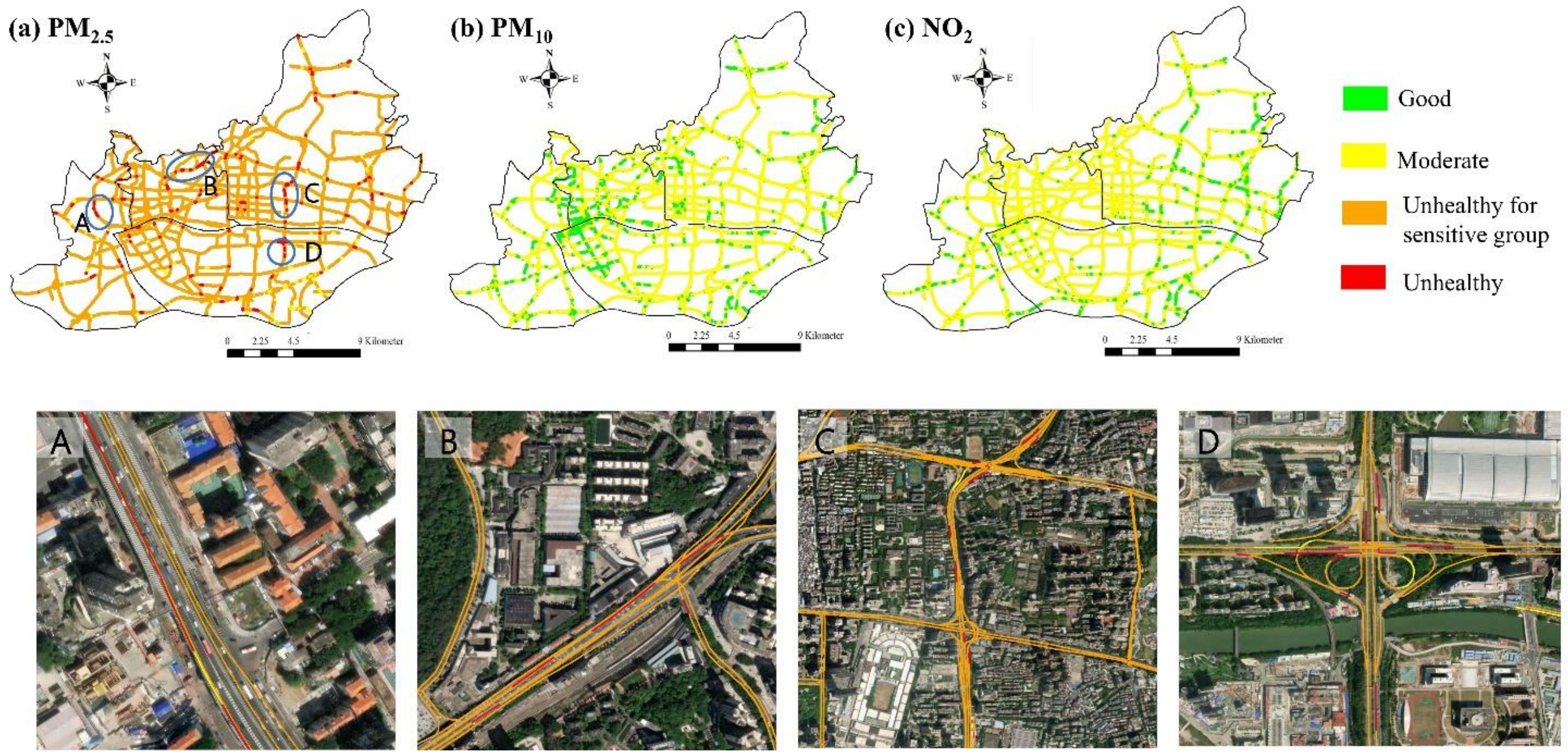

Fig.7 Spatial distribution of exposure levels for $NO_2$, $PM_{2.5}$ and $PM_{10}$

Fig. 7 illustrates the spatial distribution of exposure levels for $PM_{2.5}$, $PM_{10}$, and $NO_2$, along with high-exposure hotspots. The average exposure concentrations were recorded as 56.6 ppb for $NO_2$, 43.67 μg/m$^3$ for $PM_{2.5}$, and 56.76 μg/m$^3$ for $PM_{10}$. Regarding $NO_2$ and $PM_{10}$, their exposure levels within urban TMEs are generally not critical, with most road environments in Guangzhou's central urban area classified as Good or Moderate. Comparatively, $PM_{2.5}$ exposure represents the most significant concern, as over 68% of the road network is categorized as Unhealthy for Sensitive Groups or Unhealthy. Additionally, the $PM_{2.5}/PM_{10}$ ratio has reached 77%, further highlighting the disproportionate contribution of fine particulate matter to health risks. This suggests that vulnerable individuals, such as those with respiratory and cardiovascular conditions, face an elevated risk of exacerbation from exposure to these environments. High-exposure hotspots, including areas near schools, shopping malls, factories, and overpasses, present significant risks even to the general population. Studies have shown that short-term exposure to traffic-related particulate matter can worsen pre-existing cardiovascular and respiratory conditions, leading to symptoms such as increased heart rate, myocardial ischemia, reduced expiratory flow, and lung inflammation. These findings underscore the critical need to identify high-exposure hotspots for targeted health risk assessments and more effective urban air quality management.

## 4. Conclusions

This study utilized a taxi-based monitoring system comprising 314 vehicles to measure urban on-road air pollution exposure at a high spatial resolution of 50 meters. It evaluated the discrepancies between stationary monitoring stations and mobile monitoring systems, analyzed the spatiotemporal distribution of exposure concentrations, and classified risk levels to identify high-exposure hotspots.

The findings revealed a moderate to strong correlation between stationary and on-road measurements, with R of 0.78, 0.88, and 0.48 for $PM_{2.5}$, $PM_{10}$, and $NO_2$, respectively. However, significant discrepancies were identified between stationary and on-road measurements. Within a 500-m circular buffer under similar activity levels and environmental conditions, stationary monitoring data underestimated exposure levels, with errors of at least 35% for $PM_{10}$, up to 80% for $PM_{2.5}$, and over 100% for $NO_2$. Regarding temporal variation, on-road air pollution concentrations were significantly influenced by traffic emissions and diurnal patterns. $PM_{2.5}$ and $PM_{10}$ exhibited bimodal peaks during the morning (8:00–10:00) and evening (18:00–20:00) rush hours. $NO_2$ levels, on the other hand, were affected by solar radiation and temperature, with concentrations decreasing in the morning, reaching a trough around noon, and peaking at 18:00 due to reduced solar radiation and lower temperatures. Spatially, the average exposure concentrations across the study area were 56.6 ppb for $NO_2$, 43.67 $\mu g/m^3$ for $PM_{2.5}$, and 56.76 $\mu g/m^3$ for $PM_{10}$. While $NO_2$ and $PM_{10}$ exposure levels in most urban traffic microenvironments (TMEs) were classified as Good or Moderate, $PM_{2.5}$ emerged as the primary pollutant posing health risks. Over 68% of the road network exhibited $PM_{2.5}$ exposure levels categorized as Unhealthy for Sensitive Groups or Unhealthy, with the $PM_{2.5}/PM_{10}$ ratio reaching 77%. By leveraging high-resolution mobile monitoring data, policymakers can better design interventions to mitigate on-road air pollution exposure and protect public health, particularly for sensitive populations.

## 5. Acknowledgments

This study is funded by the National Natural Science Foundation of China under (No.52302379); Guangdong Provincial Natural Science Foundation-General Project (No.2024A1515011790); Nansha District Key Research and Development Project (No.2023ZD006); and Guangzhou Municipal Ecological Environment Bureau (No. K22-76160-026).

## Appendix A. Supplementary Material

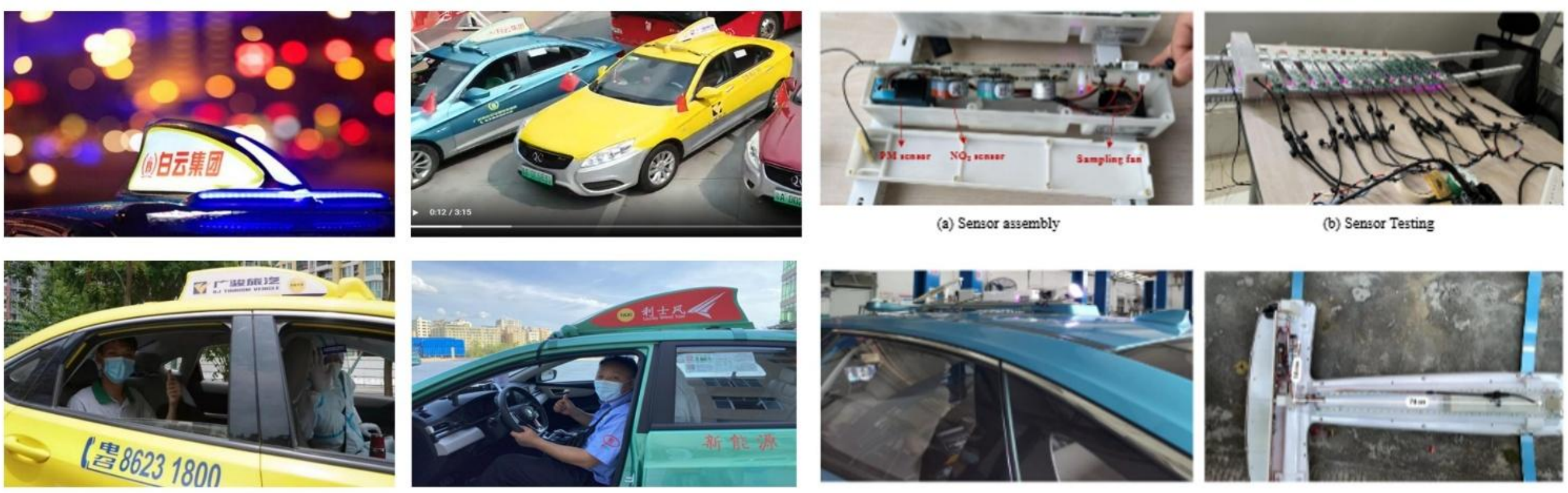

Fig.S1 Taxi-based mobile monitoring instruments

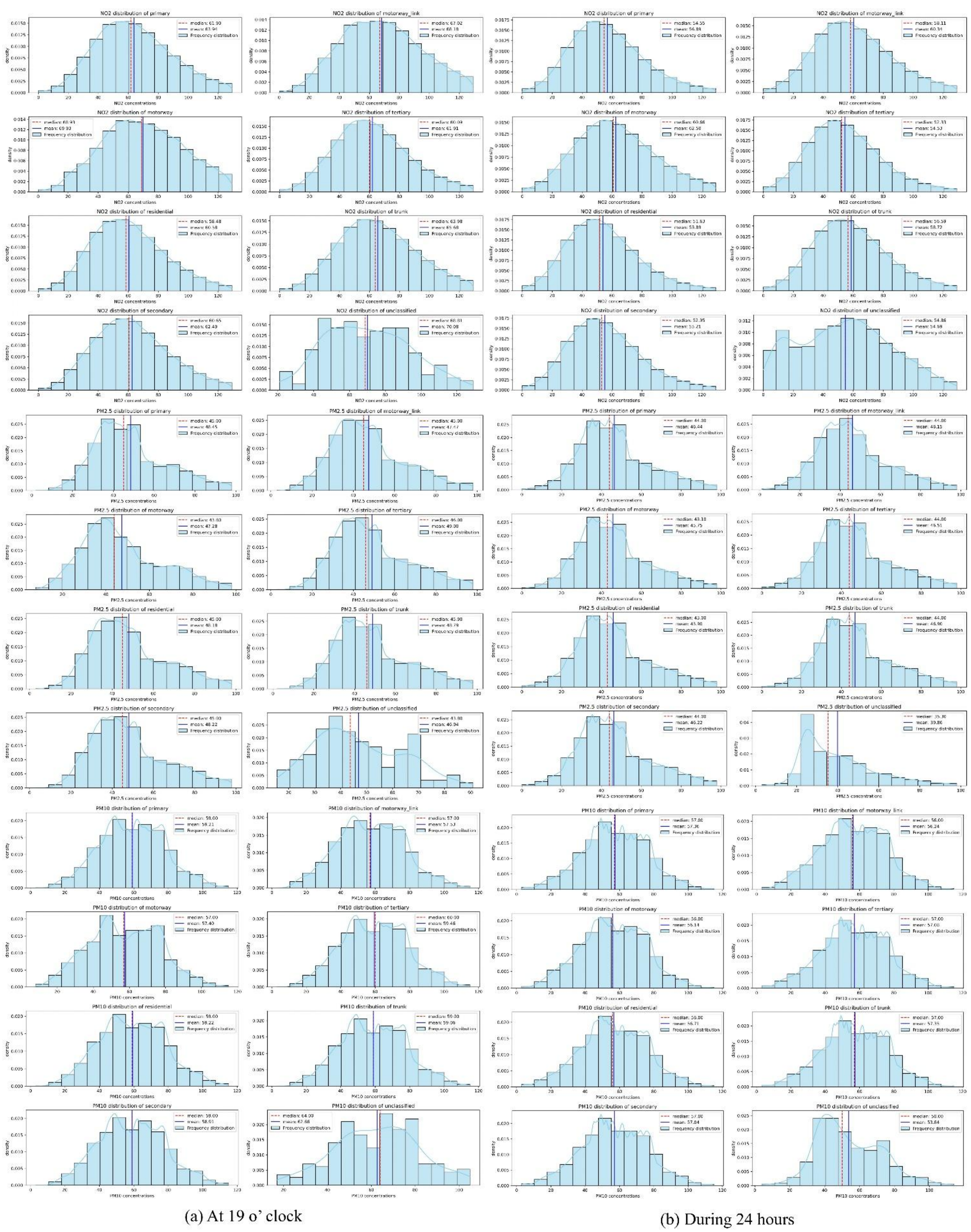

(a) At 19 o' clock

(b) During 24 hours

Fig.S2 Difference of median and mean value for NO2, PM2.5 and PM10 concentrations on various road type

Table S1 Pollutant exposure concentrations corresponding to exposure risk level

| Level | NO2 (ppb) | PM2.5 ($\mu g/m3$) | PM10 ($\mu g/m3$) |
|---|---|---|---|
| Good | 0-50 | 0-9.0 | 0-54 |
| Moderate | 51-100 | 9.1-35.4 | 55-154 |
| Unhealthy for sensitive group | 101-150 | 35.5-55.4 | 155-254 |
| Unhealthy | 151-200 | 55.5-125.4 | 255-354 |

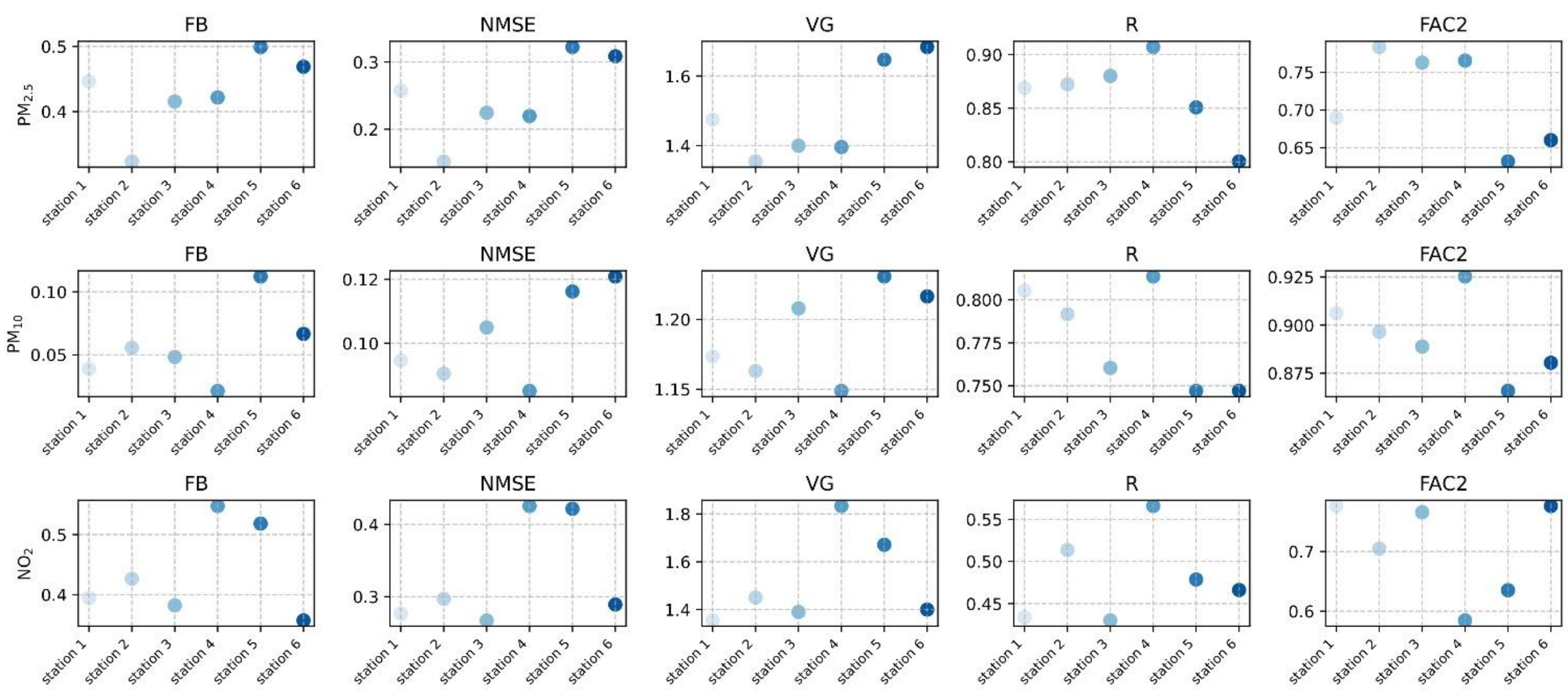

Fig.S3 Evaluation metrics of NO2, PM2.5 and PM10 exposure concentrations.

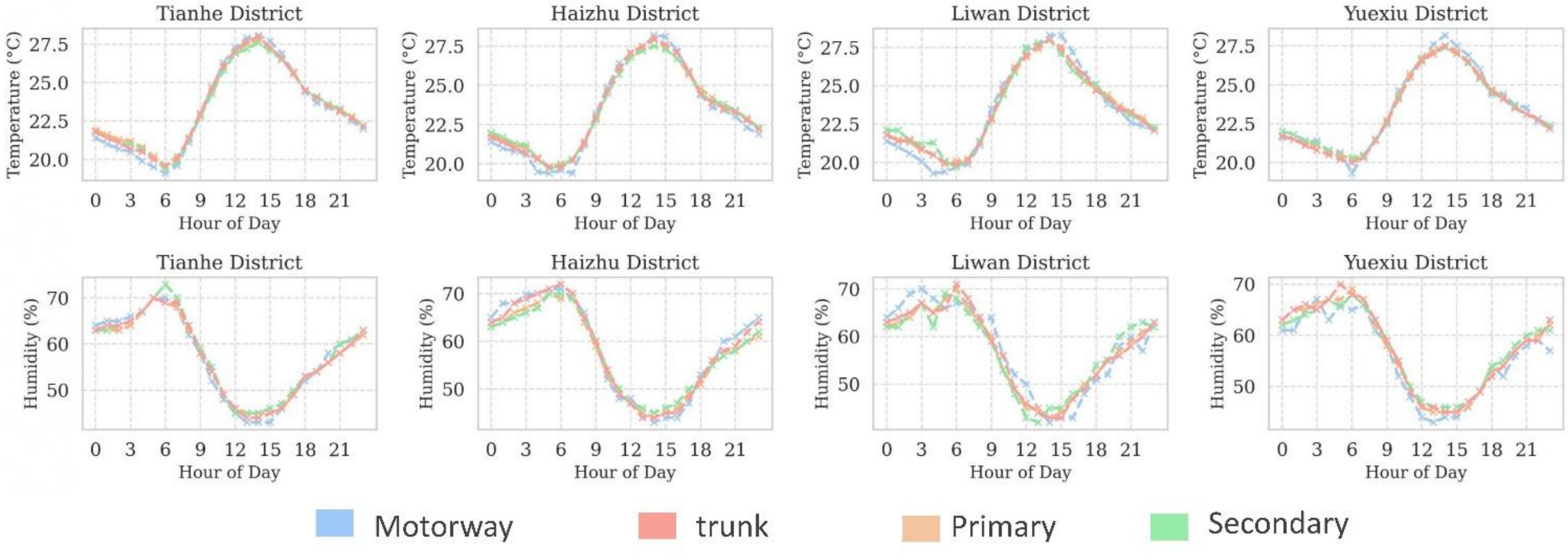

Fig.S4 Hourly variations of humidity and temperature on different road types